\magnification =\magstep 1
\hsize = 14.5 truecm
\vsize = 23 truecm
\overfullrule 0pt
\hoffset 1truecm
\baselineskip=24pt
\centerline{\bf Implications of Spontaneous Glitches in the}
\centerline{\bf Mass and Angular Momentum in Kerr Space--Time}
\vskip 4truepc
\centerline{C. Barrab\`es$^{\dagger}$$^{\ddagger}$ 
and G. F.
Bressange$^{\dagger}$\footnote*{barrabes@celfi.phys.univ-tours.fr;
bressang@celfi.phys.univ-tours.fr}}
\centerline{$^{\dagger}$ Physics Departement, UPRES A 6083 du CNRS,}
\centerline{Universit\'e de Tours, 37200 France;}
\centerline{$^{\ddagger}$ D\'epartment d'Astrophysique Relativiste et 
Cosmologie,}
\centerline{UPR 176 du CNRS, Observatoire de Paris,}
\centerline{92190 Meudon, France}
\vskip 1truepc
\centerline{and}
\vskip 1truepc
\centerline{P. A. Hogan\footnote{$^{\ast \ast}$}{phogan@ollamh.ucd.ie}}
\centerline{Mathematical Physics Department,}
\centerline{University College Dublin,}
\centerline{Belfield, Dublin 4, Ireland}
\vskip 2truepc
The outward--pointing principal null direction of the 
Schwarzschild Riemann tensor is null hypersurface--forming. 
If the Schwarzschild mass spontaneously jumps across 
one such hypersurface then the hypersurface is the 
history of an outgoing light--like shell. The outward--
pointing principal null direction of the Kerr Riemann 
tensor is asymptotically (in the neighbourhood of future 
null infinity) null hypersurface--forming. If the Kerr 
parameters of mass and angular momentum spontaneously 
jump across one such asymptotic hypersurface 
then the asymptotic hypersurface is 
shown to be the history of an outgoing light--like 
shell and a wire singularity--free spherical 
impulsive gravitational wave. 
\vfill\eject
\noindent
{\bf 1. Introduction}
\vskip 1truepc
In a recent paper [1] the authors studied the 
physical consequences of abrupt changes occuring 
spontaneously in the multipole moments of a static 
axially symmetric isolated gravitating body. The 
conclusion was that a disturbance propagates with 
the speed of light away from the source which when 
analysed near future null infinity is shown to consist 
of a spherical outgoing light--like shell accompanied 
by a spherical impulsive gravitational wave. Motivated 
by the well--known phenomenon of glitches observed 
in pulsars [2] we examine in this paper the physical 
implications of glitches in the mass and angular 
momentum associated with the source of the Kerr 
space--time. We find that in the neighbourhood of 
future null infinity a disturbance consisting of 
a spherical light--like shell and a spherical 
impulsive gravitational wave can be identified. If 
the Kerr angular momentum vanishes (the Schwarzschild 
special case) then the gravitational wave does not 
exist. If the angular momentum glitch includes an 
abrupt change in the direction of the angular momentum 
then the gravitational wave has the maximum two degrees of 
freedom of polarisation. The spherical impulsive 
gravitational waves appearing in this paper (section 3 
below) and in [1] are the only examples of such waves 
known to the authors which are free of unphysical 
directional (or wire) singularities (see section 1 of 
[1] where this is discussed).
\vskip 1truepc
To study the physical properties of the disturbances 
mentioned above we use the Barrab\`es--Israel (BI) 
theory of light--like shells and 
impulsive waves [3]. This theory is easily accessible 
and a useful description of part of it is also available in 
[1] where in particular the identification of the gravitational 
wave and the light--like shell, when both exist, is given 
explicitly. To re--derive the results stated in the present paper the 
reader must be familiar with the 
BI theory. For readers who wish to work through 
the calculations based on the BI theory we give some intermediate 
steps in the Appendix. The consequences of such calculations 
can be easily followed independently however in the 
main body of this paper.
\vskip 1truepc
To simplify the presentation and introduce our approach 
we present first in section 2 the well--known (see [3], 
for example) Schwarzschild example in which the mass 
of the source spontaneously undergoes an abrupt but 
finite change. This is also useful as a special case 
of the corresponding Kerr example , which is the main 
point of the paper, given in section 3. This is 
followed by a brief discussion of our results in 
section 4.
\vskip 4truepc
\noindent
{\bf 2. The Schwarzschild Example}
\vskip 1truepc
Consider Schwarzschild space--time with line--element 
$$ds^2=-{2r^2\,d\zeta\,d\bar\zeta\over \left (1+{1\over
2}\zeta\,\bar\zeta\right )^2}
+2du\,dr+
\left (1-{2m\over r}\right )\,du^2\ .\eqno(2.1)$$
Here $u={\rm constant}$ are future--directed null hypersurfaces 
(null--cones) generated by the geodesic integral curves of the 
null vector field $\partial/\partial r$. This vector field is 
also the outward--pointing principal null direction of the 
Riemann tensor of the space--time. We wish to consider this 
space--time undergoing a spontaneous abrupt change in the 
mass $m$ of the source across one of the outward null hypersurfaces 
$u=0$ (say) and then ask: what are the physical properties of 
$u=0$? To do this we imagine the space--time divided into two 
halves $M^+$ corresponding to $u>0$ and $M^-$ corresponding 
to $u<0$ both with boundary $u=0$ and then re--attaching the 
halves on $u=0$ preserving , with the identity map, 
the induced line--element on $u=0$:
$$dl^2=-r^2(d\theta ^2+\sin ^2\theta\,d\phi ^2)\ .\eqno(2.2)$$
We denote the resulting space--time by $M^-\cup M^+$. For the
space--time $M^-\cup M^+$ 
described above there is a stress--energy tensor concentrated 
on $u=0$ of the form 
$$T^{\mu\nu}=S^{\mu\nu}\,\delta (u)\ ,\eqno(2.3)$$
with $x^\mu =(\theta , \phi , r, u)$ and $\delta$ is the Dirac delta 
function. We refer to $S^{\mu\nu}$ as the surface stress--energy tensor 
of the light--like shell with history $u=0$ (see [3]). The 
normal to $u=0$ is the null vector with components $n^\mu$ 
given via the 1--form
$$n_\mu\,dx^\mu=du\ .\eqno(2.4)$$
The BI theory [3] gives
$$16\pi\,S^{\mu\nu}=-{4[m]\over r^2}\,n^\mu\,n^\nu\ ,\eqno(2.5)$$
where $[m]$ is the finite jump in the mass $m$ across $u=0$. 
This means that there is no stress in the out--going light--like 
shell (as might be expected because the shell is spherical 
and expanding) and the surface energy--density of the shell measured by
a 
radially moving observer (discussed in [3]) is a positive multiple 
of 
$$\sigma =-{[m]\over 4\pi\,r^2}\ ,\eqno(1.6)$$
and so it is natural to assume that $[m]<0$ for an outgoing shell. 
Thus we conclude that the space--time 
$M^-\cup M^+$ describes a Schwarzschild gravitational field (described 
by the space--time $M^-$) with an expanding 
spherical light--like shell propagating through it leaving 
behind a Schwarzschild field described by $M^+$ and with mass 
reduced compared to that of $M^-$.
\vskip 1truepc
In general in the type of situation described here (subdivision 
and re--attachment, or "cut and paste", of a space--time on a 
null hypersurface) the space--time $M^-\cup M^+$ has a 
Weyl conformal curvature tensor containing a delta function term 
singular on the null hypersurface and composed of a matter part 
(which is non--zero provided the stress in the shell is anisotropic) 
and a part describing an impulsive gravitational wave (see [1] where 
this is explicitly demonstrated and [3] for the calculation of these 
terms). However on account of 
the spherical symmetry of the Schwarzschild example above the 
Weyl tensor of $M^-\cup M^+$ vanishes identically in this case. 
Thus in particular the shell above is unaccompanied by an 
impulsive gravitational wave. 
\vfill\eject
\noindent
{\bf 3. The Kerr Example}
\vskip 1truepc
We consider here an analogous situation in the Kerr space--time 
to that considered in the Schwarzschild space--time in the 
previous section. We begin with a form of the Kerr space--time 
which makes it easy to identify the outgoing principal 
null direction of the Riemann tensor and which specialises 
to (2.1) when the Kerr angular momentum parameter is put to 
zero. In addition it will be interesting not only to consider 
spontaneous changes in the magnitude of the Kerr angular 
momentum but also to include spontaneous changes in the direction 
of the angular momentum. We thus want to use a form of the 
Kerr solution which involves the mass parameter and three 
components of the angular momentum per unit mass. One such 
form can readily be obtained by first noting that the Kerr 
solution with mass $m$ and angular momentum per unit mass $A$ 
may be written in Kerr's [4] original coordinates $(\zeta , 
\bar\zeta , r, u)$ [with the simple replacement, as in 
(2.1), of the polar angles $(\theta , \phi )$ with the 
complex coordinate $\zeta =\sqrt {2}\,{\rm e}^{i\phi}\tan\theta 
/2$ and its complex conjugate $\bar\zeta$] in the form 
$$ds^2=-2\,{\left (r^2+P^2\right )\over\left (1+{1\over 2}\zeta
\bar\zeta\right )^2}\,d\zeta\,d\bar\zeta\,+2\,d\Sigma\,\left
(dr-iP_{\zeta}\,d\zeta +
iP_{\bar\zeta}\,d\bar\zeta +S\,d\Sigma\,\right )\ ,\eqno(3.1)$$
where
$$P=A\left ({1-{1\over 2}\zeta\bar\zeta\over 1+{1\over 2}\zeta
\bar\zeta}\right )\ ,\qquad S={1\over 2}-{mr\over r^2+P^2}\
,\eqno(3.2)$$
and the 1--form $d\Sigma$ is given by
$$d\Sigma =du+iP_{\zeta}d\zeta -iP_{\bar\zeta}d\bar\zeta\ ,\eqno(3.3)$$
with $P_{\zeta}=\partial P/\partial\zeta$. The rotation
$$\zeta\rightarrow {\sqrt {2}\,\sin {\theta _1\over 2}-\zeta\,
\cos {\theta _1\over 2}\over {\rm e}^{i\phi _1}\,\cos 
{\theta _1\over 2}+{\zeta\over \sqrt {2}}\,\sin {\theta _1\over 2}}
\ ,\eqno(3.4)$$
where $\theta _1, \phi _1$ are constants, leaves the form of (3.1) 
invariant with $P$ replaced by 
$$P={a\over\sqrt {2}}\left ({\zeta +\bar\zeta\over 1+{1\over 2}
\zeta\bar\zeta}\right )+{b\over i\sqrt{2}}\left ({\zeta -\bar\zeta
\over 1+{1\over 2}\zeta\bar\zeta}\right )+c\left ({1-{1\over 2}\zeta
\bar\zeta\over 1+{1\over 2}\zeta\bar\zeta}\right )\ ,\eqno(3.5)$$
where
$$a=A\,\sin\theta _1\,\cos\phi _1\ ,\qquad b=A\,\sin\theta _1\,
\sin\phi _1\ ,\qquad c=A\,\cos\theta _1\ .\eqno(3.6)$$
We thus obtain the Kerr solution with mass $m$ and angular 
momentum 3--vector ${\bf J}=(ma, mb, mc)$ having the same magnitude
but a different direction than the initial one ${\bf J}=(0,0,mA)$. 
Restoring the polar coordinates $(\theta , \phi )$ as above we arrive at 
the line--element [5]
$$ds^2=-\rho ^2\,(d\theta ^2+\sin ^2\theta\,d\phi ^2)+2\,d\Sigma\,(dr-
N\,d\theta -M\,\sin\theta\,d\phi +S\,d\Sigma )\ ,\eqno(3.7)$$
with 
$$\eqalignno{\rho ^2&=r^2+P^2\ ,\qquad P=(a\,\cos\phi +b\,\sin\phi )\,
\sin\theta +c\,\cos\theta\ ,&(3.8)\cr
d\Sigma &=du +N\,d\theta +M\,\sin\theta\,d\phi\ ,\qquad 
S={1\over 2}-{mr\over \rho ^2}\ ,&(3.9)\cr}$$
and
$$\eqalignno{N&=-a\,\sin\phi +b\,\cos\phi\ ,&(3.10a)\cr
M&=-(a\,\cos\phi +b\,\sin\phi )\,\cos\theta +c\,\sin
\theta\ .&(3.10b)\cr}$$
We note from (3.10) that
$$E^2\equiv M^2+N^2=m^{-2}\left (\big |{\bf J}\big |^2
-\left ({\bf n}\cdot {\bf J}\right )^2\right )\ ,\eqno(
3.11)$$
where the unit 3--vector {\bf n} is given by 
${\bf n}=(\sin\theta\,\cos\phi , \sin\theta\,\sin\phi , 
\cos\theta )$. Also the outward--pointing principal 
null direction of the Riemann tensor is tangent to 
the vector field $\partial/\partial r$ or equivalently 
is given via the (non--exact) 1--form $d\Sigma$ in (3.9).
When this line element is written in a Kerr-Schild form in which
the flat background is expressed in rectangular cartesian coordinates 
and time it can be shown - see [5] in which this is described in detail-
that in the linear approximation it takes the form of the line
element of the spacetime outside the history of a slowly rotating
sphere of mass $m$ and angular momentum
${\bf J}= (ma,mb,mc)$ in exactly the same form as Kerr showed -see [4]-
that his original form approximates the line element of the spacetime 
outside the history of such a sphere of mass $m$ and angular momentum
${\bf J}= (0,0,mA)$.   
\vskip 1truepc
By analogy with the Schwarzschild example in section 2 
we might expect that a spontaneous abrupt change in the 
parameters $\{m, a, b, c\}$ will result in a disturbance 
propagating through space--time along the out--going 
principal null direction of the Kerr Riemann tensor. As 
this vector field is not surface--forming for all values 
of $r$ we cannot use the BI theory to study the 
disturbance for all $r$. However for large $r$, specifically 
if $O\left (r^{-2}\right )$--terms are neglected, then $\partial/
\partial r$ is tangent to null hypersurfaces $u={\rm constant}$. The 
normal $n^\mu$ to $u={\rm constant}$, given via the 1--form 
$n_\mu\,dx^\mu=du$, satisfies $g_{\mu\nu}\,n^\mu\,n^\nu =O\left (
r^{-2}\right )$. In this approximation $u={\rm constant}$ are portions
of future null--cones 
having the integral curves of $\partial/\partial r$ as 
generators with $r$ an affine parameter along them. These 
generators are, in the approximation under consideration, 
shear--free null geodesics with expansion $r^{-1}$ and the induced 
line--element on $u={\rm constant}$ is given 
approximately by
$$dl^2=-r^2(d\theta ^2+\sin ^2\theta\,d\phi ^2)\ .\eqno(3.12)$$
This follows from (3.7) in which the ratio of the neglected terms 
to the retained terms in the induced line--element is $O\left
(r^{-2}\right )$. 
Thus if the disturbance is propagating in the direction of $\partial/
\partial r$ then for sufficiently large values of $r$ a front is 
formed (a null hypersurface is formed in space--time) with history 
$u=0$ (say). We now assume that {\it across the null portion} of 
$u=0$ a spontaneous jump in the parameters $\{m, a, b, c\}$ occurs 
from values $\{m, a, b, c\}$ to the past $(u<0)$ of this null portion 
of $u=0$ to values $\{m_+, a_+, b_+, c_+\}$ to the future $(u>0)$ of
this 
null portion of $u=0$, and that the regions of space--time, $M^+(u>0)$
and 
$M^-(u<0)$, on either side of the null part of $u=0$ join on 
this null part with the identity map and thus preserving the 
induced line--element (3.12). We can now use the 
BI theory to study the physical properties of the null part of 
$u=0$ by calculating the surface stress--energy tensor there and by 
calculating the delta function term (the coefficient of $\delta (u)$) 
in the Weyl conformal curvature tensor.
     The results of our calculations for this Kerr example will 
naturally be a generalisation of those for the Schwarzschild 
example in section 2. Again there is a stress--energy tensor of the 
form (2.3) on the null part of $u=0$ with surface stress--energy 
described by the tensor $S^{\mu\nu}$ which in this case has 
components:
$$\eqalignno{16\pi\,S^{13}&={\left [N\right ]\over r^3}+O\left (r^{
-4}\right )\ ,&(3.13a)\cr
16\pi\,S^{23}&={\left [M\right ]\,\csc\theta\over r^3}+O\left (
r^{-4}\right )\ ,&(3.13b)\cr
16\pi\,S^{33}&=-{4[m]\over r^2}+{2\left [E^2\right ]\over r^3}+
O\left (r^{-4}\right )\ ,&(3.13c)\cr}$$
with all other components small of order $r^{-5}$. Here as before 
the square brackets denote the jump across the null part of 
$u=0$ of the quantities contained therein. $N, M, E^2$ are given 
by (3.10) and (3.11) and jump because the Kerr parameters jump. 
By (3.13) the null part of $u=0$ is the history of an outgoing 
light--like shell. By (3.13a, b) there is an anisotropic stress 
in the shell (on account of the jump in the Kerr angular momentum 
per unit mass). By (3.13c) the surface energy density of the 
shell measured by a radially moving observer is a positive 
multiple of 
$$\sigma =-{1\over 4\pi\,r^2}\left ([m]-{\left [E^2\right ]
\over 2r}+O\left (r^{-2}\right )\right )\ .\eqno(3.14)$$ 
This is the generalisation of (2.6) and $\sigma >0$ implies $[m]<0$ once
again. It is interesting 
to note that an expanding light--like shell sandwiched between 
two Reissner--Nordstrom space--times with different masses and 
charges (the charged version of the Schwarzschild example in 
section 2) has an exact surface energy density given by (3.14), 
without the error term, with $\left [E^2\right ]$ replaced by 
$\left [e^2\right ]$, the jump in the square of the charge 
across the history of the shell.
\vskip 1truepc
The BI theory enables us to calculate the coefficient of $\delta (u)$ 
in the Weyl conformal curvature tensor for the re--attached space--
time. This coefficient in 
general splits [1, 3] into a matter part, which is present if there is 
anisotropic stress in the shell (as there is in the Kerr example), 
and a wave part describing an impulsive gravitational wave 
accompanying the light--like shell. To display the components of 
this coefficient it is convenient to introduce the asymptotically 
null tetrad given via the 1--forms $du, dr+S\,du$ (with $S$ 
given in (3.9)), $\left (\sqrt{2}\right )^{-1}r\,(d\theta +i\sin\theta\,
d\phi )$ and its complex conjugate. This tetrad is asymptotically 
parallel transported along the integral curves of $\partial/\partial r$. 
By this we mean that the components on the tetrad of the covariant 
derivatives of the tetrad vectors in the direction of $\partial/
\partial r$ are small of order $r^{-2}$. Denoting the Newman--Penrose 
components on this tetrad of the matter part of the coefficient of 
$\delta (u)$ by ${}^M\Psi _A\ (A=0, 1, 2, 3, 4)$ and those of the 
wave part of this coefficient of $\delta (u)$ by ${}^W\Psi _A$, 
we find for the matter part -see [1]
$$\eqalignno{{}^M\Psi _0&=O\left (r^{-5}\right )\ ,\qquad 
{}^M\Psi _1=O\left (r^{-4}\right )\ ,\qquad 
{}^M\Psi _2=O\left (r^{-3}\right )\ ,&(3.15a)\cr
{}^M\Psi _3&=-{1\over 4\sqrt {2}r^2}\left [N-iM\right ]\,+\,O\left
(r^{-3}\right )\ ,&(3.15b)\cr
{}^M\Psi _4&=O\left (r^{-3}\right )\ ,&(3.15c)\cr}$$
and for the wave part all Newman--Penrose components vanish 
with the exception of 
$${}^W\Psi _4={1\over 4r^4}\left [m\left (N-iM\right )^2\right
]\,+\,O\left (r^{-5}\right )\ .
\eqno(3.16)$$
Here again the square brackets denote the jump across the null 
part of $u=0$ of the quantity contained therein.
\vskip 4truepc
\noindent
{\bf 4. Discussion}
\vskip 1truepc
We first notice that ${}^M\Psi _A$ is predominantly Type III in 
the Petrov classification with $n^\mu$ as degenerate principal 
null direction. The presence of ${}^M\Psi _A$ is due to the presence of 
anisotropic stress in the light--like shell (see (3.13a, b)) which 
is a consequence of the non--vanishing Kerr angular momentum parameters 
in this case. ${}^W\Psi _A$ is Type N in the Petrov classification 
with $n^\mu$ as four--fold degenerate principal null direction. This 
means that {\it the shell is accompanied by a spherical impulsive 
gravitational wave} whose presence is again due to the non--vanishing 
Kerr angular momentum parameters. Since both $N$ and $M$ are smooth 
bounded functions of $\theta , \phi$ for $0\leq\theta\leq\pi , 
0\leq\phi <2\pi$ neither (3.15) nor (3.16) possess line--singularities. 
\vskip 1truepc
It is interesting to note that the predominant radial dependence of 
${}^M\Psi _A$ and ${}^W\Psi _A$ (which is $O\left (r^{-2}\right )$ 
for ${}^M\Psi _A$ and $O\left (r^{-4}\right )$ for ${}^W\Psi _A$) is 
the same for ${}^W\Psi_A$ as in our earlier paper [1]. This is because 
the light--like shell and the gravitational wave share the same null 
hypersurface history $u=0$ in space--time and are therefore in direct 
competition with each other. Hence it is no surprise that the 
matter part is more dominant than the wave part.
\vskip 1truepc
Finally we see that if the angular momentum 3--vector {\bf J} introduced 
after (3.6) had for $u<0\ , {\bf J}=(0, 0, mc)$ and for $u>0\ ,
 {\bf J}=(0, 0, m_+c_+)$ then $N^+=N=0$ and the spherical impulsive wave
with 
amplitude (3.16) has one degree of freedom of polarisation. Adding a 
change of direction to this change of magnitude of the angular 
momentum clearly adds the extra degree of freedom to the gravitational 
wave.
\vskip 4truepc
\noindent
{\bf References}
\vskip 1truepc
\noindent
\item{[1]} C. Barrab\`es, G. F. Bressange and P. A. Hogan, {\it Phys.
Rev.} D{\bf 55}, 3477 (1997).
\vskip 1truepc
\noindent
\item{[2]} C. R. Kitchin, {\it Stars, Nebulae and Interstellar Medium} 
(Adam Hilger, Boston 1987),p.147.
\vskip 1truepc
\noindent
\item{[3]} C. Barrab\`es and W. Israel, {\it Phys. Rev.} D{\bf 43}, 1129 
(1991).
\vskip 1truepc
\noindent
\item{[4]} R. P. Kerr, {\it Phys. Rev. Lett.} {\bf 11}, 237 (1963).
\vskip 1truepc
\noindent
\item{[5]} P. A. Hogan, {\it Phys. Lett.} A{\bf 60}, 161 (1977).
\vfill\eject
\noindent
\centerline{\bf Appendix}
\vskip 1truepc
\centerline{Useful Formulas for Sections 2 and 3}
\vskip 2truepc 
\noindent
For readers who are familiar with the BI theory and 
wish to derive the results stated in sections 2 and 
3 we list here the results of some useful 
intermediate calculations. These apply to the Kerr 
example of section 3. They all specialise to the 
Schwarzschild example of section 2 when the Kerr 
angular momentum parameters are put to zero.
\vskip 1truepc
The jump $\gamma _{\mu\nu}$ in the transverse 
extrinsic curvature across the null part of $u=0$ 
is given by
$$\eqalignno{\gamma_{11}&=-2[m]+{\left [E^2\right ]\over r}+
O\left (r^{-2}\right )\ ,&(1)\cr
\gamma _{22}&=\gamma _{11}\,\sin ^2\theta\ ,&(2)\cr
\gamma _{12}&=-{\left [mMN\right ]\over r^2}\,
\sin\theta +O\left (r^{-3}\right )\ ,&(3)\cr
\gamma _{13}&=-{\left [N\right ]\over r}+O\left (
r^{-2}\right )\ ,&(4)\cr
\gamma _{23}&=-{\left [M\right ]\over r}\,\sin\theta 
+O\left (r^{-2}\right )\ ,&(5)\cr
\gamma _{33}&\equiv 0\ ,\qquad \gamma _{\mu 4}\equiv 0\ .&(6)\cr}
$$
Aswell as the $O\left (r^{-2}\right )$ leading 
term in $\gamma _{12}$ given in (3) we require the 
$O\left (r^{-2}\right )$ leading term in $\gamma _{11}
-\gamma _{22}\,\csc ^2\theta$. This is neatly given 
along with (3) by 
$$\gamma_{11}-\gamma _{22}\,\csc ^2\theta -
2\,\gamma _{12}\,\csc\theta =-{1\over r^2}\,\left [
m\left (N-iM\right )^2\right ]+O\left (r^{-3}\right )\ .\eqno(7)$$
The stress--energy tensor $S^{\mu\nu}$ of the shell 
in terms of $\gamma _{\mu\nu}$ is [3]
$$16\pi\,\eta ^{-1}S^{\mu\nu}=2\gamma ^{(\mu}\,
n^{\nu )}-\gamma\,n^{\mu}\,n^{\nu}-\gamma ^{\dagger}\,
g^{\mu\nu}-q^{\mu\nu}\ ,\eqno(8)$$
where in the present case $\eta =1+O\left (r^{-4}\right )$,
$$\gamma ^\mu =\gamma ^{\mu\nu}\,n_{\nu}\ ,\qquad 
\gamma ^{\dagger}=\gamma ^{\mu}\,n_{\mu}\ ,\qquad 
\gamma =g^{\mu\nu}\,\gamma _{\mu\nu}\ ,\eqno(9)$$
and $$q^{\mu\nu}=\epsilon\,(\gamma ^{\mu\nu}-\gamma\,g^{\mu\nu})
\ ,\eqno(10)$$
with $\epsilon =g_{\mu\nu}\,n^\mu\,n^\nu$. In the 
Kerr case $\epsilon =O\left (r^{-2}\right )$ ($\epsilon =0$ 
in the Schwarzschild case) and
$$\eqalignno{q^{11}&=O\left (r^{-6}\right )\ ,\qquad 
q^{12}=O\left (r^{-7}\right )\ ,\qquad q^{22}
=O\left (r^{-6}\right )\ ,&(11)\cr
q^{13}&=O\left (r^{-5}\right )\ ,\qquad q^{23}=
O\left (r^{-5}\right )\ ,\qquad q^{33}=O\left (
r^{-4}\right )\ ,&(12)\cr}$$
so that the $q^{\mu\nu}$ term in (8) is absorbed into the 
$O\left (r^{-4}\right )$ error in (3.13). This ensures that 
the accuracy given in (3.13) is the optimum consistent with 
$u=0$ being approximately null (in the sense that $\epsilon =
O\left (r^{-2}\right )$). 
\vskip 1truepc
The delta function term in the Weyl tensor is calculated 
from [3]
$$C^{\kappa\lambda}{}_{\mu\nu}=\left\{2\eta\,n^{[\kappa}\,
\gamma ^{\lambda ]}_{[\mu}\,n_{\nu ]}-16\pi\,\delta ^{[\kappa}
_{[\mu}\,S^{\lambda ]}_{\nu ]}+{8\pi\over 3}\,S^\alpha _\alpha\,
\delta^{\kappa\lambda}_{\mu\nu}\right\}\,\delta (u)\ .\eqno(13)$$
Care must be taken in identifying the wave part of the coefficient 
of $\delta (u)$ here. It is {\it not} given by the first term in (13) 
but is {\it contained} in the first term (the reader must consult [1] 
to see this clearly).

\bye